\documentclass[12pt]{iopart}
\usepackage{iopams}
\usepackage{amssymb}
\begin{document}
\title[Dirac's scalar field as dark energy and evolution of cosmological ``constant'']
{Dirac's scalar field as an effective component of the dark energy and an evolution of the cosmological ``constant''}
\author{O. V. Babourova, B. N. Frolov and R. S. Kostkin}
\address{Moscow State Pedagogical University, Moscow, Russia.}
\ead{\mailto{baburova@orc.ru}, \mailto{frolovbn@orc.ru}, \mailto{kostkin@orc.ru}}
\begin{abstract}
The equations of the conformal field theory of gravitation with the Dirac's scalar field in Weyl--Cartan 
spacetime have been received. Exact solutions for Dirac's scalar field for an early Universe have been derived. Intensive decrease of physical vacuum energy (dark energy) is obtained with the Dirac's scalar field as the effective cosmological ``constant''.
\end{abstract}
 
\maketitle
\section{Introduction}

It was expressed in papers \cite{BFKRusGrav13}, \cite {BFKIzvVuz2008} that a gauge approach to the theory of gravitation should be added to the existing geometrical approach. As it is well known, a gauge approach allows to construct the modern classical field theory. It was shown in \cite{BFZ1}--\cite{BFZ3} on the base of the gauge theory of the Poincar\'{e}--Weyl group that the Weyl--Cartan spacetime geometry is the background to the modern theory of gravitation.

The Poincar\'{e}--Weyl group is a nontrivial (noncommutative) union of space-time dilatations (stretching and compression) and the Poincar\'{e} group. Dilatations are mathematically equivalent to the group of changes of a scale, which is the basis of the gauge theory proposed by Weyl in 1918 \cite{Weyl2}.

In \cite{BFZ1}--\cite{BFZ3} the gauge theory for the Poincar\'{e}--Weyl group has been constructed. 
The actuality of this approach follows from the fact that high-energy physics requires the local scale 
invariance. This theory is based on the method of introduction of gauge fields for groups 
associated with the transformations of spacetime coordinates. This method is founded on the first and the 
second Noether theorems and developed in \cite{Fr1}--\cite{FrIv}. Note that 
in this approach tetrad coefficients $h^a{}_\mu$ are not gauge fields but the functions of the true gauge fields.

The gauge field, which introduced by a subgroup of dilatations, is called dilatation field. Its potential vector is the Weyl vector, and the tension is the segmental curvature tensor appearing together with the curvature and torsion tensors after the geometric interpretation of the theory. A special feature of the gravitational field Lagrangian, which has been built in \cite{BFZ1}--\cite{BFZ3}, is the existence of a nonzero mass for the Weyl's nonmetricity vector field without violation of the gauge invariance. This means that the dilatation gauge field does not constitute an electromagnetic field (which is claimed by Weyl in his basic work \cite{Weyl2}), but a field of another type \cite{Ut2}--\cite{Ut3}.

Note also that in the papers \cite{BFZ1}--\cite{BFZ3} in the frames of gauge procedure the Dirac's scalar field $\beta(x)$  \cite{Dir} has been naturally introduced. This field plays a crucial role in the construction of the proposed gravitational field Lagrangian. Some terms in this Lagrangian have the structure of the Higgs Lagrangian, and thus can cause the spontaneous breaking of dilatation invariance, which leads to the creation of particle masses \cite{Frolov4}.

It is well known that a conformal symmetry (in particular the Weyl scale symmetry) is of a great importance in the quantum field theory. Breaking of this symmetry at the quantum level is connected with the determination of the structure of counterterms, with the problem of asymptotic freedom in the quantum field theory, as well as with the calculation of the critical dimensions ($n = 26$ and $n = 10$) in the string theory, with gravitational instantons, with the phenomenon of Hawking evaporation black holes, with the problems of inflation, with the cosmological constant, with the particle creation and creation black holes in the early universe \cite{Duff}. Construction of the conformal theory of physical fields at the quantum level is currently one of the most urgent problems of fundamental physics.

An important part of this problem is the creation of an adequate conformal classical field 
theory, in particular, the conformal theory of the gravitational field. A solution to this problem can be the gauge theory of Poincar\'{e}--Weyl group \cite{BFZ1}--\cite{BFZ3}, in which a conformally invariant Lagrangian of the gravitational field has been constructed. In the geometric interpretation of this theory, a curved space is appearing with the tangent space, in which the metric tensor has the form $g^{gauge}_{ab}=\beta^2(x)g^M_{ab}$, where $g^M_{ab}$ is the metric tensor of Minkowski space. Thus this tangent space is no longer Minkowski space, but a Minkowski--Weyl space. This type of metric tensor is used by Strominger in his famous heterotic string theory \cite{Stromg}.

However in the affine-metric theory of gravitation, it is considered the tangent space to be the Minkowski 
space. In order to superpose both of these theories one can to redefine components of the metric tensor in the coordinate space: $g_{\mu\nu} = \beta^{-2}g^{gauge}_{\mu\nu}$. It leads to a redefinition of other geometric quantities of the theory. Resulting from this procedure is a conformal theory of gravity in the Weyl--Cartan spacetime 
\cite{BFKGRACOS2009}--\cite{BFK_Izv} with an additional geometric structure of the Dirac's scalar field  $\beta$, which in this approach arises naturally as a necessary element of the theory. In other theories, which dealt with the scalar field in affine-metric theory of gravity \cite{Gre-Pap1}, \cite{Gre-Pap2}, this field is introduced into the theory from the outside ``by hands'', which is rather artificial.

The scaling-invariant theory of gravitation with scalar field in a Riemann space-time has
been developed in \cite{Perv1}--\cite{Arbuzov} (see also the references therein) in order to derive an alternative scenario of the evolution of the Universe. In this scenario some of the observation data of
modern cosmology can be explained without introducing the $\Lambda$-term and without adopting the inflation hypothesis. In contradiction with this, we do not reject the inflation and the $\Lambda$-term.
In \cite{BFKGRACOS2009}--\cite{BFK_Izv} we have expressed the hypothesis that the dark energy is determined by the value of the Dirac scalar field via the term $\Lambda_0 \beta^4$ of the Lagrangian. The conformal theory with a scalar field in the Weyl--Cartan spacetime that is developed in the paper proposed is wider than the theory developed in \cite{Perv1}--\cite{Arbuzov} because it contains (in comparison with that) two additional geometrical quantities -- the Weyl vector and the torsion tensor. This alters significantly the variational equations of the gravitational theory.

In the second part of this paper the conformal transformations induced by the localized Poincar\'{e}--Weyl group are introduced. In the third part the variational formalism of the conformal theory of gravity in the Weyl--Cartan spacetime with the Dirac's scalar field is developed. As a result, there are three variational equations of the gravitational field: results of the variation of the connection ($\Gamma$-equation), the tetrads ($h$-equation) and the Dirac's scalar field ($\beta$-equation). In the fourth part of this paper, these equations are used to solve the problem of changing of the cosmological constant and the problem of the energy of physical vacuum, which is determined by the cosmological constant. In the fifth part the differential identities are obtained that can be used to check the validity of the variational derivatives and to found some correlations between the coupling constants of the gravitational Lagrangian. 

\section{Conformal transformations induced by the localized Poincar\'{e}--Weyl group}

In the Poincar\'{e}--Weyl gauge theory of gravitation \cite{BFZ1}--\cite{BFZ3}, the gauge invariant tensions of the gauge fields (Lorentz $r$-field, translational $t$-field and dilatational $d$-field) are the geometric quantities of the Weyl--Cartan spacetime (expressed in the tetrad form): 
a curvature tensor $R^{a}{}_{b\mu\nu}$, a torsion tensor $T^a{}_{\mu\nu}$, a  segmental curvature tensor $V_{\mu\nu}$ and a Weyl's nonmetricity vector $Q_\mu $, 
\begin{eqnarray}
R^{a}{}_{b\mu\nu}= 2\partial_{[\mu}{\Gamma^a{}_{|b|\nu]}}+2\Gamma^a{}_{c[\mu}\Gamma^c{}_{|b|\nu]},
\nonumber\\
T^a{}_{\mu\nu}=2\partial_{[\mu}h^a{}_{\nu]}+2\Gamma^a{}_{b[\mu}h^b_{\nu]},
\nonumber\\
V_{\mu\nu} = \nabla_{[\mu}Q_{\nu]} + \frac{1}{2}T^\lambda{}_{\mu\nu}Q_\lambda\,,
\nonumber\\
Q_\mu = g_{\alpha\beta}Q^{\alpha\beta}{}_\mu\;,\quad Q^{\alpha\beta}{}_\mu =\nabla_\mu g^{\alpha\beta} =  \frac{1}{4} g^{\alpha\beta}Q_{\mu}  \;.
\label{eq:Cond}
\end{eqnarray}

In this theory, the tangent space is not a Minkowski space, but a Minkowski--Weyl space. Its metric tensor in a special coordinate system can be represented as follows:
\begin{eqnarray*}
g^{gauge}_{ab} = \beta^2(x)g^M_{ab}
\end{eqnarray*}
where $g^M_{ab}$ is the metric tensor of a Minkowski space, and $\beta$ is the Dirac's scalar field. 

Under localized infinitesimal dilatation transformations in the tangent space with the parameter $\varepsilon(x)$, the field $\beta$ and the tetrads transform as follows:
\begin{eqnarray}
\delta\beta = \varepsilon(x)\beta\, ,\qquad \delta h^a{}_\mu = -\varepsilon h^a{}_\mu\, ,
\label{eq:preobofbeta}
\end{eqnarray}
but the Minkowski metric $g^M_{ab}$ is not transformed. Therefore, in this approach the metric tensor of the coordinate space $g_{\mu\nu}$ is also not transformed according to the formula
\begin{eqnarray}
g_{\mu\nu}= g_{ab}h^a{}_\mu h^b{}_\nu\, ,
\label{eq:gcoorbaz}
\end{eqnarray}
where $g_{ab}=\beta^2(x)g^M_{ab}$. 

However, in the conventional treatment of a Weyl--Cartan spacetime it is considered the tangent space to be the Minkowski space, the metric tensor of which, as already indicated, does not transform under the transformations of the dilation. But according to (\ref{eq:gcoorbaz}), where in this case 
$g_{ab}=g^M_{ab}$, the metric tensor $g_{\mu\nu}$ of coordinate space should transform under the transformations of dilatation:
\begin{eqnarray}
\delta g_{ab} = 0\,, \qquad \delta g_{\mu\nu} = -2\varepsilon g_{\mu\nu}\,.
\label{eq:gminkgaugcoor}
\end{eqnarray}

We will follow to this standard treatment of dilatation transformations based on the formulas (\ref{eq:preobofbeta}) and (\ref{eq:gminkgaugcoor}). The geometric quantities will thus be transformed as follows:
\begin{eqnarray}
&\delta \Gamma^a{}_{b\mu} = \delta^a_b\partial_\mu\varepsilon\, ,\quad \delta Q^{ab}{}_\mu = 2g^{ab}\partial_\mu\varepsilon\,, \quad \delta  Q_\mu = 8\partial_\mu\varepsilon\,,
\label{eq:gaugpreob1}\\
&\delta R^a{}_{b\mu\nu} = 0\, ,\quad \delta T^a{}_{\mu\nu} = -\varepsilon T^a{}_{\mu\nu}\,,\quad \delta V_{\mu\nu} = 0\,.
\label{eq:gaugpreob2}
\end{eqnarray}

\section{Variational procedure in the tetrad conformal theory of gravity with a scalar field in a  Weyl--Cartan spacetime}

A variational procedure in a Weyl--Cartan spacetime can be done by various methods: independent variation of a metric, torsion and a Weyl vector; independent variation of tetrads, Lorentz connection and a Weyl vector \cite{MinkGarKud}; a metric and a general holonomic connection of $L_4(g,\Gamma)$ space using the condition (\ref{eq:Cond}) with the help of Lagrange multipliers \cite{IntJ}, \cite{ModPhysLet}, \cite{Fr:book}.

In this paper we generalize  variational formalism developed in \cite{BKU}, \cite{BabK:izv} to the case of the presence of the Dirac's scalar field. The field equations are obtained with the help of the tetrad formalism in the theory of gravitation with quadratic Lagrangians using the variational first-order formalism, in which metric and nonholonomic connection are treated as independent variational variables (generalized Palatini formalism, see \cite{Schr}--\cite{Fr3}). The condition (\ref{eq:Cond}) is taking into account using Lagrange multipliers. This formalism has been used erlier in \cite{BabK:izv} without the Dirac's scalar field to be taken into consideration. 

In the presence of the Dirac's scalar field $\beta$, the Lagrangian density of the gravitational field,  which is invariant under conformal transformations of (\ref{eq:preobofbeta}), (\ref{eq:gminkgaugcoor})--(\ref{eq:gaugpreob2},) is the following:
\begin{eqnarray} 
\mathcal{L}_G= \sqrt{-g}(f_0\beta^2 R + L_{R^2}+\beta^2 L_{T^2} +\beta^2 L_{Q^2} + \beta^2 L_{TQ} + L_\beta) \, ,
\label{eq:LG2}
\end{eqnarray}
where
\begin{eqnarray}
\fl
L_{R^2} =
f_1R^{{\alpha}{\beta}{\mu}{\nu}}R_{{\alpha}{\beta}
{\mu}{\nu}}+f_2R^{{\alpha}{\beta}{\mu}{\nu}}R_{{\beta}{\alpha}{\mu}{\nu}}+f_3R^{{\alpha}{\beta}{\mu}{\nu}}R_{{\alpha}{\mu}{\beta}{\nu}}+
f_4R^{{\alpha}{\beta}{\mu}{\nu}}R_{{\beta}{\mu}{\alpha}{\nu}}\nonumber
\\
+f_5R^{{\alpha}{\beta}{\mu}{\nu}}R_{{\mu}{\nu}{\alpha}{\beta}}+
f_6R^2+f_7R^{{\mu}{\nu}}R_{{\mu}{\nu}}+f_8R^{{\mu}{\nu}}R_{{\nu}{\mu}}+ f_9R^{{\mu}{\nu}}{\tilde R}_{{\mu}{\nu}}\nonumber
\\
+f_{10}R^{{\mu}{\nu}}{\tilde R}_{{\nu}{\mu}}+ f_{11}{\tilde R}^{{\mu}{\nu}}{\tilde R}_{{\mu}{\nu}}+ f_{12}{\tilde R}^{{\mu}{\nu}}{\tilde R}_{{\nu}{\mu}}+ f_{13}V^{{\mu}{\nu}}R_{{\mu}{\nu}}\nonumber\\
f_{14}V^{{\mu}{\nu}}{\tilde R}_{{\mu}{\nu}}+ f_{15}V^{{\mu}{\nu}}V_{{\mu}{\nu}}
\label{eq:LR2}
\end{eqnarray}
is the Lagrangian quadratic in curvature,
\begin{eqnarray}
L_{T^2} = a_1T^{{\lambda}{\mu}{\nu}}T_{{\lambda}{\mu}{\nu}}+
a_2T^{{\lambda}{\mu}{\nu}}T_{{\nu}{\mu}{\lambda}}+ a_3T^\mu T_\mu
\label{eq:LT}
\end{eqnarray}
is the Lagrangian quadratic in torsion,
\begin{eqnarray}
\fl
L_{Q^2} = k_1Q^{{\mu}{\nu}{\lambda}}Q_{{\mu}{\nu}{\lambda}}+
k_2Q^{{\mu}{\nu}{\lambda}}Q_{{\mu}{\lambda}{\nu}}+k_3Q^\mu Q_\mu + 
k_4Q_{\lambda}{}^{\mu}{}_{\mu}Q^{{\lambda}{\nu}}{}_{\nu}+k_5Q^\mu Q_{\mu}{}^\nu{}_{\nu}
\label{eq:LQ}
\end{eqnarray}
is the Lagrangian quadratic in nonmetricity,
\begin{eqnarray}
L_{QT} = m_1Q^{{\mu}{\nu}{\lambda}}T_{{\mu}{\nu}{\lambda}}+
m_2Q^{\mu}T_{\mu}+ m_3Q^{\mu}{}_{{\nu}{\mu}}T^{\nu}
\label{eq:LQT}
\end{eqnarray}
is the Lagrangian containing contractions nonmetricity and torsion, 
\begin{eqnarray}
\fl
L_{\beta}  = l_1g^{\mu\nu}\partial_\mu\beta\partial_\nu\beta + l_2\beta\partial_\mu\beta g^{\mu\sigma} T_\sigma + l_3\beta\partial_\mu\beta g^{\mu\sigma}Q_\sigma + l_4\beta \partial_\mu\beta Q^{\mu\sigma}{}_{\sigma} + \Lambda_0 \beta^4
\label{eq:LBeta}
\end{eqnarray}
is the Dirac's scalar field Lagrangian.

The full Lagrangian density of the theory is as follows:
\begin{eqnarray}
\mathcal{L} = \mathcal{L}_G + \mathcal{L}_m + \frac12\sqrt{-g}\beta^4\Lambda^\mu{}_{ab}\left(Q^{ab}{}_\mu -\frac14 g^{ab}Q_\mu\right)\, ,\quad \Lambda^\mu{}_{ab}g^{ab} = 0\, ,
\label{eq:Lagr}
\end{eqnarray}
where $\mathcal{L}_m $ is the Lagrangian density of the sources of the gravitational field, and the last term is the additional term with the Lagrange multiplier $\Lambda^\mu{}_{ab}$. This term ensures fulfillment of the Weyl condition (\ref{eq:Cond}) as a result of the variational procedure.

In carrying out the variational procedure, it is effective to use the following formulas:
\begin{eqnarray}
\fl
\sqrt{-g}\beta^2H^{\nu\mu b}{}_a \delta R^a{}_{b\mu\nu} = \partial_\nu (2\sqrt{-g}\beta^2H^{[\mu\nu ]b}{}_a \delta \Gamma^a{}_{b\mu}) \nonumber\\
+\left (2\stackrel{*}{\nabla}_\nu (\sqrt{-g}\beta^2H^{[\nu\mu ]b}{}_a ) -\sqrt{-g}\beta^2H^{\alpha\beta b}{}_a T^\mu{}_{\alpha\beta}\right )
\delta \Gamma^a{}_{b\mu}\;,\label{eq:newdR}
\end{eqnarray}
\begin{eqnarray}
\fl
\sqrt{-g}\beta^2N_a{}^{\mu\nu}\delta T^a{}_{\mu\nu} =\partial_\nu (2\sqrt{-g}\beta^2N_a{}^{[\nu\mu ]} \delta h^a{}_\mu ) -
2\sqrt{-g}\beta^2N_a{}^{[b\mu ]}\delta \Gamma^a{}_{b\mu}\nonumber\\
+ \left (2\stackrel{*}{\nabla}_\nu (\sqrt{-g}\beta^2H_a{}^{[\mu \nu]} ) +
\sqrt{-g}\beta^2\,T^{\mu\sigma\nu}H_{a\sigma\nu}\right )\delta h^a{}_\mu \;, \label{eq:newdT}
\end{eqnarray}
\begin{eqnarray}
\fl
\sqrt{-g}\beta^2B_{ab}{}^\mu \delta Q^{ab}{}_\mu = \partial_\mu (\sqrt{-g}\beta^2B_{(ab)}{}^\mu \delta g^{ab}) \nonumber\\
 - \stackrel{*}{\nabla}_\mu (\sqrt{-g}\beta^2B_{(ab)}{}^\mu )\delta g^{ab} +
2\sqrt{-g}\beta^2B_{(a}{}^{b)\mu }\delta \Gamma^a{}_{b\mu} \;. \label{eq:newddQ}
\end{eqnarray}

The variational derivatives of the Lagrangian density (\ref{eq:LG2}) are the following: with respect to the connection,
\begin{eqnarray}
\fl
\frac{\delta\mathcal{L}_G}{\delta\Gamma^{a}{}_{b\lambda}} = f_0 \frac{\delta(\beta^2\sqrt{-g} R)}{\delta\Gamma^{a}{}_{b\lambda}} + \frac{\delta(\sqrt{-g}L_{R^2})}{\delta\Gamma^{a}{}_{b\lambda}} +\frac{\delta(\beta^2\sqrt{-g} L_{T^2})}{\delta\Gamma^{a}{}_{b\lambda}} + \frac{\delta(\beta^2\sqrt{-g}L_{Q^2})}{\delta\Gamma^{a}{}_{b\lambda}}\nonumber\\
+\frac{\delta(\beta^2\sqrt{-g}L_{QT})}{\delta\Gamma^{a}{}_{b\lambda}} + \frac{\delta(\sqrt{-g} L_{\beta})}{\delta\Gamma^{a}{}_{b\lambda}}\; ,
\label{eq:VarG}
\end{eqnarray}
with respect to the tetrads, 
\begin{eqnarray*}\
\fl
\frac{\delta\mathcal{L}_G}{\delta h^{a}{}_{\mu}} = f_0 \frac{\delta(\sqrt{-g}\beta^2R)}{\delta h^{a}{}_{\mu}} +  \frac{\delta(\sqrt{-g} L_{R^2})}{\delta h^{a}{}_{\mu}} + \frac{\delta(\sqrt{-g} \beta^2L_{T^2})}{\delta h^{a}{}_{\mu}} + \frac{\delta(\sqrt{-g}\beta^2L_{Q^2})}{\delta h^{a}{}_{\mu}} \nonumber\\
+\frac{\delta(\sqrt{-g}\beta^2 L_{QT})}{\delta h^{a}{}_{\mu}} + \frac{\delta(\sqrt{-g}L_{\beta})}{\delta h^{a}{}_{\mu}}\;,
\end{eqnarray*}
and with respect to the scalar field, 
\begin{eqnarray*}
\fl
\frac{\delta\mathcal{L}_G}{\delta \beta} = 2\beta \sqrt{-g}\left(f_0 R+L_{T^2} + L_{Q^2} + L_{QT}\right) + \frac{\delta{(\sqrt{-g}L_\beta})}{\delta\beta}\;.
\end{eqnarray*}
By varying the total Lagrangian density (\ref{eq:Lagr}) with respect to the Lagrange multipliers $\Lambda^\mu{}_{ab}$ one obtains the Weyl condition (\ref{eq:Cond}) on the nonmetricity tensor, 
\begin{eqnarray}
Q^{ab}{}_\mu = \frac14g^{ab}Q_\mu\, ,
\label{eq:Cond2}
\end{eqnarray}
which should be taken into account in the resulting variational field equations.

The resulting variational field equations one can see in \ref{appA}. These equations will be used for obtaining and solving the equation for the Dirac's scalar field in the early Universe. It should be noted that since the Lagrangian density (\ref{eq:Lagr}), prior to the procedure of variation, has been recorded in the general affine-metric space, and only after the variation procedure one can see that spacetime is the Weyl--Cartan spacetime, then we can formally write down the result of variation of this Lagrangian density under the components of the metric tensor $g^{ab}$. One can prove that these equations are a consequence of the rest field equations of the theory (see Sec. 5). 

\section{The solution of the equation for the Dirac's scalar field at the early stage of the Universe evolution}

By antisymmetrizating and calculating the trace, the variational $\Gamma$-equation (\ref{eq:vGbeta}) splits into two independent equations.  After taking into account the Weyl conditions (\ref{eq:Cond}) and the following definitions for the torsion trace and the modified torsion tensor, respectively,  
\begin{eqnarray*}
&&T_\mu = T^\mu{}_{ab}h^b{}_\mu\;, \qquad M^\mu{}_{ab} = T^\mu{}_{ab} + 2h^\mu{}_{[a}T_{b]}\;,
\end{eqnarray*}
the first of these equations reads,
\begin{eqnarray}
\fl
f_0\sqrt{-g}\beta^2(M^\mu{}_{ab} - \frac12h^\mu{}_{[a}Q_{b]}) + (2\delta^{\mu}_{[\alpha}\delta^{\nu}_{\beta]}\nabla_\nu + M^\mu{}_{\alpha\beta} +
\frac12 h^\mu{}_{[\alpha}Q_{\beta]})\sqrt{-g}(2(f_1 \nonumber\\
- f_2)R_{[ab]}{}^{\alpha\beta}-(2f_3 -f_4)R_{[a}{}^{\alpha\beta}{}_{b]} + f_4R^{\alpha}{}_{[ab]}{}^{\beta} +2f_5R^{\alpha\beta}{}_{ab} \nonumber
\\+
2f_6Rh^\alpha{}_{[a}h^\beta{}_{b]}+ (2f_7 + f_9)R_{[a}{}^{\alpha}h^\beta{}_{b]} +
(2f_8 + f_{10})h^\alpha{}_{[a}R^\beta{}_{b]} \nonumber
\\+
(f_9 + 2f_{11})\tilde{R}_{[a}{}^{\alpha}h^\beta{}_{b]} +(f_{10} + 2f_{12})h^\alpha{}_{[a}\tilde{R}^{\beta}{}_{b]}+(f_{13} + f_{14})V_{[a}{}^{\alpha}h^{\beta}{}_{b]})\nonumber
\\+
\sqrt{-g}\beta^2(4a_1T_{[ab]}{}^{\mu} - 2a_2(T^\mu{}_{ab} + T_{[ab]}{}^\mu)
 +2a_3h^\mu{}_{[a}T_{b]})\nonumber
\\ - \frac14\sqrt{-g}\beta^2(m_1-4m_2 -m_3)h^\mu{}_{[a}Q_{b]} \nonumber\\
+\sqrt{-g}\beta^2\partial_\nu\ln\beta (-4f_0+l_2)h^\mu{}_{[a}h^\nu{}_{b]}= -\sqrt{-g}S^\mu{}_{ab} \;.  
\label{eq:grur1}
\end{eqnarray}
The second of these equations reads,
\begin{eqnarray}
\nabla_\nu\biggl(\sqrt{-g}\Bigl(V^{\mu\nu}(4f_1 + 4f_2 + 2f_{13} - 2f_{14} + 16f_{15}) 
\nonumber\\+
R^{[\mu\nu]}(4f_3 + 2f_4 + 4f_7 - 4f_8 - 2f_9 + 2f_{10} +8f_{13}) 
\nonumber\\+
\tilde{R}^{[\mu\nu]}(-2f_4 + 2f_9 - 2f_{10} -4f_{11} + 4f_{12} +8 f_{14})\Bigr)\biggr)
\nonumber\\+
\sqrt{-g}\biggl(V^{\alpha\beta}(2f_1 + 2f_2 + f_{13} - f_{14} + 8f_{15})
\nonumber\\
+ R^{\alpha\beta}(2f_3 + f_4 + 2f_7 - 2f_8 - f_9 +f_{10} +4f_{13})
\nonumber\\
+ \tilde{R}^{\alpha\beta}(-f_4 + f_9 - f_{10} - 2f_{11} + 2f_{12} +4f_{14})\biggr)M^\mu{}_{\alpha\beta}
\nonumber\\
+\sqrt{-g}\beta^2\biggl[T^\mu(4a_1 + 2a_2 +6a_3 -2m_1 + 8m_2 + 2m_3)
\nonumber\\+
Q^\mu\Bigl(4k_1 + k_2 + 16k_3 +k_4 + 4k_5 - \frac34(m_1 + 4m_2 + m_3)\Bigr) 
\nonumber\\
+g^{\mu\nu}\partial_\nu\ln\beta(3l_2 + 8l_3 +2l_4)\biggr] = -\sqrt{-g}J^\mu \;. 
\label{eq:grur2}
\end{eqnarray}

We shall consider that period of the early Universe, in which the birth of elementary particles with nonzero rest masses (as a consequence of spontaneuos breaking of the dilatational invariance) has not yet happened, and the decisive contribution to the dynamics of the Universe gives a Dirac's scalar  field. For this reason, we neglect in the field equations terms with material sources. In addition, at this stage of the study we will ignore terms in the Lagrangian of the gravitational field depending on the squares of the curvature tensor ($f_i = 0\,,\; i=1,\,2,\,\dots,\,15$).

With these asumptions, in the antisymmetric part of $\Gamma$-field equation, Eq. (\ref{eq:grur1}), we equate to zero its source: $S^\mu{}_{ab} = 0 $, and form the contraction of this equation with tetrad $h^b{}_\mu $. The result can be summarized as follows:
\begin{eqnarray}
t_1 T_\mu + q_1 Q_\mu + b_1 \partial_\mu\ln\beta  =0\,,
\label{eq:TQ1}
\end{eqnarray}
where the coefficients are
\begin{eqnarray*}
t_1 = 2f_0+2a_1+a_2+3a_3\,,\\
q_1 = -\frac34f_0+\frac38(-m_1+4m_2+m_3)\,,\\
b_1 = \frac32(-4f_0+l_2)\,.
\end{eqnarray*}

Now let us consider the trace of $\Gamma$-field equation, Eq. (\ref{eq:grur2}), equate to zero its  source: $J^\mu = 0$, and the terms appearing from the squares of the curvature in (\ref{eq:LG2}). The resulting equation is as follows:
\begin{eqnarray}
t_2 T_\mu + q_2 Q_\mu + b_2 \partial_\mu\ln\beta =0\,,
\label{eq:TQ2}
\end{eqnarray}
where the coefficients are
\begin{eqnarray*}
t_2 = 4a_1+2a_2+6a_3-2m_1+8m_2+2m_3\,,\\
q_2 = 4k_1+k_2+16k_3+k_4+4k_5-\frac34(m_1+4m_2+m_3)\,,\\
b_2 = 3l_2+8l_3+2l_4\,.
\end{eqnarray*}

We shall consider equations (\ref{eq:TQ1}) and (\ref{eq:TQ2}) as a system of linear algebraic equations for unknown variables $T_\mu$ and $Q_\mu$. Calculating these values, we find
\begin{eqnarray}
&T_\mu = \chi_T\partial_\mu\ln\beta\,, \qquad
Q_\mu = \chi_Q\partial_\mu\ln\beta\,,
\label{eq:TQb}
\end{eqnarray}
where the coefficients are
\begin{eqnarray*}
\chi_T= \frac{q_1b_2 - q_2b_1}{t_1q_2 - t_2q_1}\,,\qquad
\chi_Q = -\frac{b_1t_2-b_2t_1}{q_1t_2-q_2t_1}\,.
\end{eqnarray*}

Let us consider the $\beta$-equation (\ref{eq:betaur}), equate to zero its source,  $\frac{\delta\mathcal{L}_m}{\delta\beta} = 0$, and the terms appearing from the squares of the curvature in (\ref{eq:LG2}). As a result we obtain: 
\begin{eqnarray*}
\fl
 - 2l_1\stackrel{*}{\nabla}_\mu\left(\sqrt{-g}g^{\mu\nu}\partial_\nu\beta\right) - \beta\stackrel{*}{\nabla}_\mu\left(\sqrt{-g}(l_2T^\mu + l_3Q^\mu + l_4Q^{\mu\lambda}{}_\lambda)\right)
\\
+ 2\beta\sqrt{-g}\biggl(f_0R + L_{T^2} + L_{Q^2} + L_{TQ}\biggr) + 4\sqrt{-g}\Lambda\beta^3 = 0\,.
\end{eqnarray*}
Substitute in this equation the Weyl condition (\ref{eq:Cond2}), multiply by $\beta$ and express from the result the term with the scalar curvature:
\begin{eqnarray}
\fl
2\beta^2\sqrt{-g}f_0R = -2\beta^2\sqrt{-g}(L_{T^2} + L_{Q^2} + L_{TQ}) + 2l_1\beta\stackrel{*}{\nabla}_\mu(\sqrt{-g}g^{\mu\nu}\partial_\nu\beta) 
\nonumber\\+
\beta^2\stackrel{*}{\nabla}_\mu(\sqrt{-g}(l_2T^\mu +(l_3 + \frac14l_4)Q^\mu)) - 4\sqrt{-g}\Lambda\beta^4 = 0\,.
\label{eq:urR}
\end{eqnarray}

Let us consider now the $h$-field equation (\ref{eq:vhbeta}), equate to zero its source,  $\frac{\delta\mathcal{L}_m}{\delta h^a{}_\mu} = 0 $, and the terms appearing from the squares of the curvature in (\ref{eq:LG2}). Then we form the contraction of this equation with tetrad $h^a{}_\mu$. The result is as follows:
\begin{eqnarray}
\fl\sqrt{-g}\beta^2\biggl[2f_0R+4(L_{T^2}+L_{Q^2}+L_{TQ})-2a_1T^{\mu\sigma\nu}T_{\mu\sigma\nu}-2a_2T^{\nu\mu\sigma}T_{\sigma\mu\nu} - 2a_3T^\mu T_\mu 
\nonumber\\
-2k_1Q^{\nu\lambda\mu}Q_{\nu\lambda\mu}-2k_2Q^{\mu\nu\lambda}Q_{\nu\lambda\mu}-2k_3Q^\mu Q_\mu-2k_4Q^{\lambda\mu}{}_{\mu}Q_\lambda{}^\nu{}_{\nu}
\nonumber\\-
k_5(Q^{\mu\nu}{}_\nu Q_\mu + Q_\nu Q^{\nu\mu}{}_\mu) - 2m_1T^{\nu\lambda\mu}Q_{\nu\lambda\mu} - 2m_2T^\mu Q_\mu - 2m_3Q^{\mu\nu}{}_\nu T_\mu\biggr] 
\nonumber\\+
\sqrt{-g}\biggl[4L_\beta - 2l_1 g^{\sigma\rho}\partial_\sigma\beta\partial_\rho\beta - 2\beta\partial_\nu\beta (l_2 T^\nu + l_3Q^\nu + l_4Q^\nu{}_\lambda{}^\lambda)\biggr]
\nonumber\\+
\stackrel{*}{\nabla}_\nu\biggl[\beta^2\sqrt{-g}(-4a_1T^\nu - 8a_2T^\nu -6a_3T^\nu + 2m_1Q_\mu{}^{[\mu\nu]} -3m_2Q^\nu
\nonumber\\+
2m_3\delta^{[\nu}_{\mu}Q^{\mu]\lambda}{}_\lambda + 2l_2\partial_\sigma\ln\beta g^{\sigma[\mu}\delta^{\nu]}_{\mu})\biggr] + 4\sqrt{-g}\Lambda^\sigma{}_{bc}Q^{bc}{}_\sigma  = 0\,.
\label{eq:vhh}
\end{eqnarray}

Since these equations are applied to the solution of the cosmological problem at the early stage of the Universe evolution, we assume homogeneity and isotropy of spacetime and of the scalar field distribution. We choose the metric in the form of the  Friedmann--Robertson--Walker (FRW) metric, 
\begin{eqnarray}
&ds^2 = dt^2 - a^2(t)\left( \frac {dr^2}{1-kr^2} + r^2(d\theta^2 + \sin^2 \theta d\phi^2 ) \right ) \,.\label{eq:RUF}
\end{eqnarray}
Then, as it was shown in \cite{Tsamp}--\cite{CQG}, torsion field is determined only by its trace:
\begin{eqnarray}
T_{\lambda\mu\nu} = -\frac13g_{\lambda\mu}T_\nu + \frac13g_{\lambda\nu}T_\mu\,.
\label{eq:TorSl}
\end{eqnarray}

Let us substitute this relation into the previous relation (\ref{eq:vhh}) and also take into account the Weyl condition (\ref{eq:Cond2}). Since  according to (\ref{eq:Lagr}) the relation $\Lambda^\mu{}_{ab}g^{ab}=0$ is valid for the Lagrange multipliers, we obtain the following equation:
\begin{eqnarray*}
\fl\sqrt{-g}\beta^2\biggl[2f_0R+4\left(L_{T^2}+L_{Q^2} + L_{TQ}\right) + T^\nu T_\nu\left(-\frac43a_1-\frac23a_2-2a_3\right)
\\+
Q^\mu Q_\mu\left(-\frac12k_1 -\frac18k_2 -2k_3 - \frac18k_4 -\frac12k_5\right)+ T^\mu Q_\mu \biggl(\frac12m_1 - 2m_2
\\-
\frac12m_3\biggr)\biggr]+ \sqrt{-g}\biggl[4L_\beta -2l_1g^{\sigma\rho}\partial_\sigma\beta\partial_\rho\beta -
\beta\partial_\nu\beta\biggl(2l_2T^\nu + \biggl(2l_3  
\\+
\frac12l_4\biggr)\biggr)Q^\nu\biggr] + \stackrel{*}{\nabla}_\nu\biggl[\beta^2\sqrt{-g}\biggl((-4a_1-8a_2-6a_3)T^\nu +
\biggl(\frac34m_1 - 3m_2 
\\-
\frac34m_3\biggr)Q^\nu - 3l_2g^{\sigma\rho}\partial_\sigma\ln\beta\biggr) \biggr ]=0\,.
\end{eqnarray*}

After substituting into this equation the expression for the scalar curvature (\ref{eq:urR}), the scalar curvature $R$ will be excluded in the resulting equation. Now let us calculate the values of Lagrangians  $L_{T^2}$, $L_{Q^2} $ and $L_{TQ}$ by substituting the relation (\ref{eq:TorSl}) and the Weyl condition (\ref{eq:Cond2}). The result of these calculations are as follows:
\begin{eqnarray*}
L_{T^2} = \frac13(a_1+a_2-a_3)T^\nu T_\nu\,,\\
L_{Q^2} = (\frac14k_1 +\frac{1}{16}k_2+k_3+\frac1{16}k_4+\frac14k_5)Q_\nu Q^\nu\,,\\
L_{TQ} = \left(-\frac14m_1 - m_2 - \frac14m_3\right)Q_\nu T^\nu\,.
\end{eqnarray*}

After some additional transformations we obtain the equation
\begin{eqnarray}
\fl\sqrt{-g}\beta^2\biggl[\left(-\frac23a_1 -\frac83a_3\right)T^\nu T_\nu +(-4m_2 -m_3)Q_\nu T^\nu \biggr]+\stackrel{*}{\nabla}_\nu\biggl[\beta^2\sqrt{-g}(-4a_1 - 8a_2-\nonumber\\-6a_3+l_2)T^\nu + \biggl(\frac34m_1 -3m_2 -\frac34m_3 +l_3 +
\frac14l_4\biggr)Q^\nu +\nonumber\\+  (2l_1 -3l_2)g^{\sigma\nu}\partial_\sigma\ln\beta\biggr] = 0\,.
\label{eq:betaur2}
\end{eqnarray}
In the equation (\ref{eq:betaur2}) we substitute the expressions for the torsion and nonmetricity traces via the scalar field (\ref{eq:TQb}): 
\begin{eqnarray}
\sqrt{-g}\beta^2 g^{\mu\nu}\partial_\mu\ln\beta\partial_\nu\ln\beta\biggl[\chi_T^2\biggl(-\frac23a_1 -\frac83a_3\biggr) + \chi_Q\chi_T(-4m_2 -m_3)\biggr] +\nonumber\\
+\stackrel{*}{\nabla}_\nu\biggl[\beta^2\sqrt{-g}g^{\mu\nu}\partial_\mu\ln\beta\biggl\{(-4a_1 -8a_2 -6a_3 +l_2)\chi_T +\nonumber\\+
\biggl(\frac34m_1 - 3m_2 - \frac34m_3 + l_3 + \frac14l_4\biggr)\chi_Q + (2l_1 -3l_2)\biggr\}\biggr]  = 0\,.
\label{eq:betaur3}
\end{eqnarray}

In this equation we calculate the term with $\stackrel{*}{\nabla}_\nu$ using the formula
\begin{eqnarray*}
\stackrel{*}{\nabla}(\sqrt{-g}w^\nu) = \partial_\nu(\sqrt{-g}w^\nu)\,.
\end{eqnarray*}
After differenivation we verify that all terms with explicit dependence on the metric tensor vanish, if we assume that the metric tensor satisfies the condition,   
\begin{eqnarray}
\partial_\nu(\sqrt{-g}g^{\mu\nu }) = 0\,.
\label{eq:garm}
\end{eqnarray}

The coordinate system, in which this condition is satisfied, is called a harmonic one. In the book \cite{Fock} it was shown that the harmonic coordinate system exists for the open Friedmann universe ($k =- 1$ in the metric (\ref{eq:RUF})). According to current cosmological observational data, the metric of the Universe at large distances is spatially flat ($k=0$ in the metric (\ref{eq:RUF})). It is easy to show that the harmonic coordinate system also exists for a spatially flat Friedmann universe, and the harmonic time (but not the spatial coordinates) coincides with the time of the FRW coordinate system. 

If one goes to the harmonic system of coordinates in a spatially flat Universe, the equation for the Dirac's scalar field (\ref{eq:betaur3}) takes the following form (for the early evolution of the Universe, when dynamics of the universe determined only by a scalar field and ordinary matter has not yet been generated):
\begin{eqnarray}
A\partial_\mu\partial_\nu\ln\beta - B\partial_\mu\ln\beta\partial_\nu\ln\beta  = 0\,,
\label{eq:betaur4}
\end{eqnarray}
where the coefficients depend on the coupling constants of the initial gravitational field Lagrangian (\ref{eq:LG2}):
\begin{eqnarray*}
A =&\chi_T(-4a_1 -8a_2 -6a_3 +l_2) + \chi_Q \biggl(\frac34m_1 -3m_2 - \frac34m_3 + l_3 +\nonumber\\&+\frac14 l_4\biggr)+ 2l_1 -3l_2\,,\\
B = &\chi_T^2\left(\frac23a_1 +\frac83a_3\right)+\chi_T\chi_Q(4m_2+m_3) - 2A\,.
\end{eqnarray*}

In a homogeneous and isotropic Universe the scalar field can depend only on time: $\beta=\beta(t)$. In this case equation (\ref{eq:betaur4}) takes the form
\begin{eqnarray}
&\beta \ddot \beta -(q+1) (\dot \beta )^2 = 0\,, \qquad q=\frac{B}{A}\,. 
\label{eq:zur}
\end{eqnarray}

Replacing $\beta=u^{(-1/q)}$ for $q\neq0$ or $\beta=\exp u$ for $q=0 $, this equation reduces to $\ddot u=0 $ . Then the solution of the equation (\ref{eq:zur}) will have the form
\begin{eqnarray}
\beta = \frac{\beta_0}{(C_1t + 1)^{(1/q)}} \,,\quad (q\neq 0), \qquad  \beta = \beta_0 \exp (-C_2 t)\,,\quad (q=0)\,, 
\label{eq:b(t)}
\end{eqnarray}
where $\beta_0 $ and $C_1$, $C_2$ are arbitrary constants of integration. The constant $\beta_0$ is the initial value of the field at the time $t=0$ i.e. at the Planck time. From considerations of quantum field theory this quantity should be very large. The constants $C_1$ and $C_2$ are determined by the initial rate of the field changing, and to reduce the field $\beta$ in time this quantity must satisfy the condition $\dot{\beta}_0 < 0$.

The parameter $q$ is determined in sufficiently complex manner by 16 of the coupling constants of the original Lagrangian (\ref{eq:LG2}): three constants in the quadratic torsion Lagrangian (\ref{eq:LT}), five constants in nonmetricity quadratic Lagrangian (\ref{eq:LQ}), three constants of the interaction Lagrangian of torsion and nonmetricity (\ref{eq:LQT}), four constants of the Lagrangian of the scalar field $\beta$ (\ref{eq:LBeta}) and one constant $f_0$ before the curvature scalar. Therefore we can always choose these constants so that one of the conditions is valid, 
\begin{eqnarray}
\frac{1}{q} = \frac{A}{B} >> 1\,, \quad \mbox{or} \quad q = 0\quad (\mbox{if}\quad B = 0).  
\label{eq:>>1}
\end{eqnarray} 

Thus (at least for an open or flat universe) if one of the conditions (\ref{eq:>>1}) is valid, one can provide the necessary rapid decline with time the value of the Dirac's scalar field. This in turn would explain the rapid (up to 120 orders during the existence of the Universe) a decrease of energy of physical vacuum (dark energy) $\Lambda_0\beta ^ 4$. The decrease of the value of dark energy by the law (\ref{eq:b(t)}) is much more intensive than the corresponding decrease, which can be carried out in Poincare gauge theory of gravity \cite{MinkGarKud}, \cite{Mink2}, \cite{FrKouch}, in which effective cosmological constant is determined by the trace of the torsion tensor.

It should be pointed out that the rapid decrease of the energy of physical vacuum be the law (\ref{eq:b(t)}) will occur only before the Friedmann era of evolution of the Universe begins, since according to the scenario of inflation  the birth of rest masses of elementary particles occurs in the late period of inflation. Further evolution of the Universe will not be determined only by Dirac's scalar field, but the ultrarelativistic matter interacting with radiation. In this case the above solution requires an appropriate modification.

\section{The differential identities}

The variational field equations (\ref{eq:vGbeta}), (\ref{eq:vhbeta}), (\ref{eq:betaur}) and  (\ref{eq:vg}) are not independent because of existing two differential identities for the  variational derivatives. This fact is established by the following theorem, in which we follows \cite{Fr:book}. 

{\em Theorem.} If in a general affine-metric space $L_{4}(g,\,\Gamma)$ the action integral for a gravitational Lagrangian density, 
\begin{equation}
{\cal L}_G = {\cal L}_G (h^a\!_\mu\,,\partial_\sigma h^a\!_\mu\,,\Gamma ^a\!_{b\mu}\,,\partial_\sigma \Gamma^a\!_{b\mu}\,,\beta\,,\partial_\sigma \beta )\,,
\label{eq:GLagr} 
\end{equation}
is $SL(4,R)$-gauge and diffeomorfic invariant, then the strong differential identities are valid between variational derivatives of (\ref{eq:GLagr}) with respect to a connection $\Gamma^a\!_{b\lambda}$, tetrads $h^a\!_\mu$, a metric tensor $g^{ab}$ and the Dirac's scalar field $\beta$,   
\begin{eqnarray}
\stackrel{*}{\nabla}_\mu \left (\frac{\delta {\cal L}_G}{\delta\Gamma
^a\!_{b\mu}}\right ) + \frac{\delta{\cal L}_G}{\delta h^a\!_\mu} h^b\!_\mu
+ 2\frac{\delta {\cal L}_G}{\delta g^{ac}} g^{cb} = 0\; , \label{eq:tjd1} \\
\stackrel{*}{\nabla}_\mu \left (\frac{\delta {\cal L}_G}{\delta h^a\!
_\mu}\right ) + \frac{\delta{\cal L}_G}{\delta h^c\!_\mu} T^c\!_{\mu a} +
\frac{\delta {\cal L}_G}{\delta\Gamma^{b}\!_{c\mu}} R^b\!_{c\mu a} -
\frac{\delta {\cal L}_G}{\delta g^{bc}} Q^{bc}\!_a - \frac{\delta {\cal L}_G}{\delta \beta} h^\mu{}_a \partial_\mu \beta= 0\; . \label{eq:tjd2}
\end{eqnarray}

{\em Proof.} Let us consider the variations of the independent variables under the infinitesimal transformations of the gauge group $SL(4,R)(x)$ acting in the tangent space of an affine-metric space $L_{4}(g,\,\Gamma)$, 
\[
\delta\Gamma^a\!_{b\lambda} = \omega^a\!_c \Gamma^c\!_{b\lambda} -
\omega^c\!_b \Gamma^a\!_{c\lambda} - \partial_\lambda\omega^a\!_b \, ,\;\,
\delta h^a\!_\mu = \omega^a\!_b h^b\!_\mu \, ,\;\, 
\delta g_{ab} = -\omega_{(ab)} \, , \;\, \delta \beta =0\, ,
\]
where $\omega^a\!_b = \omega^a\!_b(x)$ are the infinitesimal parameters of the gauge group $SL(4,R)$, which are arbital differential functions of spacetime points.  

After substituting these variations into the variation of the action integral for the Lagrangian density (\ref{eq:GLagr}), we get, 
\begin{eqnarray}
&& 0 = \int_\Omega \,(d^4x)\,\delta {\cal L}_G  \nonumber \\ && = \int_
\Omega \,(d^4x)\left [\stackrel{*}{\nabla}_\mu\left (\frac{\delta
{\cal L}_G}{\delta\Gamma^a_{b\mu}}\right ) + \frac{\delta{\cal L}_G}
{\delta h^a\!_\mu} h^b\!_\mu + 2\frac{\delta {\cal L}_G}{\delta g^{ac}} g^{cb}
\right ]\omega^a\!_b (x)  \nonumber\\ && +\int_\Omega\,(d^4x)\,\left
[-\partial_\mu\left (\frac{\delta{\cal L}_G}{\delta\Gamma^a_{b\mu}}
\omega^a\!_b \right ) +\mbox{total divergence}\right ]\;.\label{eq:tin0}
\end{eqnarray}
Here the ``total divergence'' depends on the $\omega^a\!_b(x)$, which is equal to zero on the boundary of the domain of integration $\Omega$. As the consequence of the randomnes of $\omega^a\!_b (x)$ inside the domain $\Omega$, we obtain the first of the identity (\ref{eq:tjd1}). \newline
Let us now get the consequence of the diffeomorfic invariance of the action intergral, which is equivalent to the invariance under the general coordinate transformations. In this case  the coordinate basis $\vec e_\mu  = \vec e_a h^a\!_\mu$ is transformed, but nonholonomic basis $\vec e_a$ is not transformed. Other independent variables are transformed as follows, 
\begin{eqnarray}
\delta\Gamma^a\!_{b\mu} = - \Gamma^a\!_{b\sigma}  \partial_\mu\delta
x^\sigma \;  ,  \quad  \delta  h^a\!_\mu = - h^a\!_\sigma\partial_\mu\delta
x^\sigma \; , \quad \delta g^{ab} = 0\; , \quad \delta \beta =0\; . \label{eq:txGh}
\end{eqnarray}
One introduces the function form variation, $\bar\delta = \delta - \delta x^\sigma\partial_\sigma$, which commutes with differentiation. Then the variation of the action integral reads, 
\begin{eqnarray}
&& 0 = \delta \int_\Omega \,(d^4x)\,{\cal L}_G = \int_\Omega\,(d^4x)\,
 (\mbox{total divergence})\nonumber \\ 
 && = \int_\Omega \,(d^4x)\left (\frac{\delta{\cal L}_G}{\delta\Gamma^a_{b\mu}}
\bar\delta\Gamma^a\!_{b\mu} + \frac{\delta{\cal L}_G}{\delta
h^a\!_\mu}\bar\delta h^a\!_\mu + \frac{\delta {\cal L}_G}{\delta g^{ab}}
\bar\delta g^{ab} + \frac{\delta {\cal L}_G}{\delta \beta}
\bar\delta \beta \right ) \; , \label{eq:txint0}
\end{eqnarray}
where the ``total divergence'' depends on the $\delta x^\sigma(x)$, which is equal to zero on the boundary of the domain of integration $\Omega$. As the consequence of the randomnes of $\delta x^\sigma(x)$ inside the domain $\Omega$, we obtain from (\ref{eq:txint0}) the following identity, 
\begin{eqnarray}
\partial_\mu \left (\frac{\delta{\cal L}_G}{\delta\Gamma^a_{b\mu}}
\right ) \Gamma^a\!_{b\sigma} + 2\frac{\delta{\cal L}_G}{\delta\Gamma^a_{b
\mu}} \partial_{ [\mu}\Gamma^a\!_{\vert b\vert\sigma ]} \nonumber \\
+ \partial_\mu \left (\frac{\delta{\cal L}_G}{\delta h^a\!_\mu}\right )
h^a\!_\sigma + 2\frac{\delta{\cal L}_G}{\delta h^a\!_\mu} \partial_{ [\mu}
h^a\!_{\sigma ]} - \frac{\delta{\cal L}_G}{\delta g^{ab}}
\partial_\sigma g^{ab} - \frac{\delta {\cal L}_G}{\delta \beta} \partial_\sigma \beta= 0\; . \label{eq:tjd3}
\end{eqnarray}
This identity with the help of the identity (\ref{eq:tjd1}) transforms to the gauge and diffeomorfic covariant identity (\ref{eq:tjd2}). This is the end of the proof. 

{\em Comments.} The differential identities similar to (\ref{eq:tjd1}) and (\ref{eq:tjd2}) (but without the Dirac's scalar field) in a Riemann--Cartan space, in a Weyl--Cartan space or in the general affine-metric space have been used by many autors. In a Riemann--Cartan space these identities have been established in \cite{Tr2} and they have been used in \cite{YadPhys2} in order to derive the equations of motion of a test particle with spin and color charge. In metric formalism the identity (\ref{eq:tjd2}) has been proved in \cite{BabFrKor} on the basis of Lie derivative method and in \cite{BFKIzvVuz2008} by method of general coordinate transformations. In exterior form formalism the identities (\ref{eq:tjd1}) and (\ref{eq:tjd2}) have been established in \cite{He-CMN} in the general affine-metric space and in \cite{DilSing2}, \cite{CQG} in a Weyl--Cartan space. In the tetrad formalism these identities have been proved in \cite{Fr:book}.   

The identities (\ref{eq:tjd1}) and (\ref{eq:tjd2}) can be used in order to check the validity of calculations of variational equations (\ref{eq:vGbeta}), (\ref{eq:vhbeta}), (\ref{eq:betaur}) and  (\ref{eq:vg}). For each terms of the Lagrangian densities (\ref{eq:LR2})--(\ref{eq:LBeta}) the identities (\ref{eq:tjd1}) and (\ref{eq:tjd2}) have been calculated and in all cases we have obtained results equal to zero. 

One can check the validity of these identities also with the help of simbolic calculatins on computer. With this aim we have used the system CartanWeyl \cite{BFKgrandkosm2009}--\cite{BFKPIRT2009}. This system is a modification of the well known system CARTAN \cite{Soleng} with the aim to perform simbolic calculatins on computer with the geometric quantities of Weyl--Cartan space. With the help of  CartanWeyl we have veryfied the vanishing of these differential identities for all terms of the Lagrangian densities (\ref{eq:LR2})--(\ref{eq:LBeta}). These two testing means that the variational equations (\ref{eq:vGbeta}), (\ref{eq:vhbeta}), (\ref{eq:betaur}) and (\ref{eq:vg}) have been correctly calculated. 

In the theory developed, one can obtain the third differential identity corresponding to invariance of the action integral for the Lagrangian density (\ref{eq:LG2}) with respect to the conformal transformations (\ref{eq:preobofbeta}), (\ref{eq:gminkgaugcoor})--(\ref{eq:gaugpreob2}). In order to derive this identity let us substitute these conformal transformations to the variation of the action integral for (\ref{eq:GLagr}), form a total divergence and equate the rezult to zero. As a consequence of an randomness of $\varepsilon (x)$ inside the domain of integration $\Omega$ and its vanishing on the boundary of $\Omega$, we obtain the following identity,                             
\begin{eqnarray}
& \stackrel{*}{\nabla}_\mu \left (\frac{\delta {\cal L}}{\delta\Gamma
^a\!_{b\mu}}\delta^a_b\right ) + \frac{\delta{\cal L}}{\delta h^a\!_\mu} h^a\!_\mu
- \frac{\delta {\cal L}}{\delta \beta} \beta = 0\, . \label{eq:diftoj3}
\end{eqnarray}

This identity is valid, only if some relations exist between coupling constants of the Lagrangian densities (\ref{eq:LR2})--(\ref{eq:LBeta}). After substituting to the identity (\ref{eq:diftoj3}) the variational derivatives (\ref{eq:vGbeta}), (\ref{eq:vhbeta}) and (\ref{eq:betaur}), we obtain the equation,   
\begin{eqnarray}
&&\stackrel{*}{\nabla}_\mu \biggl(\sqrt{-g}\beta^2 \Bigl(Q^\mu(4k_1+k_2+16k_3+\frac12k_4+4k_5+l_3+\frac14l_4) \nonumber\\
&&+T^\mu(-2m_1+8m_2+2m_3+l_2)+g^{\mu\nu}\partial_\nu\ln\beta(2l_1+8l_3+2l_4)\Bigr)\biggr) = 0\,.
\end{eqnarray}
This equation is an identity, if the following relations beteen the coupling constants is sutisfied, 
\begin{eqnarray}
&&l_1 = 16k_1 + 4k_2 + 64k_3 + 2k_4 + 16k_5\,,\\
&&l_2 = 2m_1 - 8m_2 - 2m_3\,, \qquad l_1 + 4l_3 + l_4 = 0\,.
\label{eq:svjaz}
\end{eqnarray}

Therefore it has been demonstrated that an investigation of the differential identities (\ref{eq:tjd1}),   (\ref{eq:tjd2}) and (\ref{eq:diftoj3}) is an effective method of the verification of the variational field equations and the discovery of some important relations beteen the coupling constants of the gravitational field Lagrangian. 

\section{Conclusion}

Construction of the conformal theory of gravity in a Weyl-Cartan spacetime with the Dirac's scalar field allows to raise a question on solving of one of the most important problems of modern fundamental physics. It refers to the problem of huge differences (up to 120 orders of magnitude) in values of the cosmological constant $\Lambda$ in the period of inflation and in the modern era \cite{CosmConPr}. 

In the gravitational field Lagrangian (\ref{eq:LG2}), which is invariant under the Poincar\'{e}--Weyl gauge group, the  term $\Lambda\beta^4 $ is interpreted as an effective cosmological constant. Therefore in the theory proposed the effective cosmological constant is determined by the Dirac's scalar field  \cite{BFK_lanl}. Solving the equation for the scalar field for the early Universe (inflation phase), a solution with sharp decrease in the value of the Dirac's scalar field in this era is obtained, which provides a sharp decrease in the effective cosmological constant to its present level over the lifetime of the Universe \cite{BFKGRACOS2009}, \cite{BFK_lanl}. This allows to make a substantial progress in solving the cosmological constant (dark energy) problem.

Thus it is shown that in conformal theory of gravity with the Dirac's scalar field and quadratic Lagrangians in a Weyl--Cartan spacetime it can be resolved one of the fundamental contradictions of the theory of evolution of the Universe \cite{CosmConPr}, \cite{Rubakov}, and the coordination of this theory with modern fundamental physical concepts can be fulfilled.
\newline

This research work has been performed in the framework of the Federal Purposeful Program "Research
and Pedagogical Personnel of Innovative Russia" for 2009-–2013.

\appendix
\section{The field equations}
\label{appA}
With the help of the formulas (\ref{eq:newdR})--(\ref{eq:newddQ}) we find the expressions for the corresponding variational derivatives with respect to the connection $\Gamma^a{}_{b\mu}$. We substitute them in (\ref{eq:VarG}), and then in the variational derivative of the total Lagrangian density theory (\ref{eq:Lagr}). As a result, we obtain the variational field equation corresponding to one of the independent variable - the connection ($\Gamma$-equation),
\begin{eqnarray}
\fl
\frac{\delta\mathcal L}{\delta {\Gamma}^a{}_{b{\mu}}} = 
2\stackrel{*}{\nabla}_{\nu}\biggl[\sqrt{-g}\Bigl(2f_1R_a{}^{b{\mu}{\nu}}+
2f_2R^b{}_a{}^{{\mu}{\nu}}+ 2f_3R_a{}^{[{\mu}|b|{\nu}]}- f_4(R^{b[\mu\nu]}{}_{a}+R^{[{\mu}}{}_{|a|}{}^{{\nu}]b})
\nonumber\\
+ 2f_5R^{[{\mu}{\nu}]}{}_a{}^b+2f_6Rh^{[{\mu}}{}_{|a|}h^{{\nu}]}{}_{c} g^{bc}+
2f_7R^{b[{\nu}}h^{{\mu}]}{}_{a}+2f_8h^{[{\mu}}{}_{|a|}R^{{\nu}]b}
\nonumber\\+
f_9({\tilde R}^{b[{\nu}}h^{{\mu}]}{}_a+R_a{}^{[{\mu}}h^{{\nu}]}{}_cg^{bc})+
f_{10}(h^{[{\mu}}{}_{|a|}{\tilde R}^{{\nu}]b}+g^{bc}h^{[{\nu}}{}_{|c|}R^{{\mu}]}{}_a)\nonumber\\
+ 2f_{11}{\tilde R}_{a}{}^{[{\mu}} h^{{\nu}]}{}_c g^{bc}+
2f_{12}g^{bc}h^{[{\nu}}{}_{|c|}{\tilde R}^{{\mu}]}{}_{a}
+ f_{13}({\delta}^b_aR^{[{\mu}{\nu}]}+  V^{b[{\nu}}h^{{\mu}]}{}_a)
\nonumber\\
+ f_{14}({\delta}^b_a{\tilde R}^{[{\mu}{\nu}]}+g^{bc}V_{a}{}^{[\mu}h^{\nu ]}{}_{c})+
2f_{15}{\delta}^b_aV^{{\mu}{\nu}}\Bigr )\biggr]-\sqrt{-g}\beta^2f_0P^{{\mu}b}{}_a
\nonumber\\
-\sqrt{-g}\Bigl(2f_1R_a{}^{b{\beta}{\alpha}}+2f_2R^b{}_a{}^{{\beta}{\alpha}}+
2f_3R_a{}^{{\alpha}{\beta}b}+f_4(R^{b{\beta}}{}_a{}^{{\alpha}}+R^{{\beta}}{}_a{}^{b{\alpha}})
\nonumber\\
+2f_5R^{{\beta}{\alpha}}{}_a{}^b+2f_6Rh^{{\alpha}}{}_ch^{{\beta}}{}_ag^{bc}+ 2f_7h^{{\beta}}{}_aR^{b{\alpha}}+2f_8h^{{\beta}}{}_aR^{{\alpha}b}
\nonumber\\
+ f_9(h^{{\beta}}{}_a{\tilde R}^{b{\alpha}}+ h^{{\alpha}}{}_c g^{bc}R_a{}^{{\beta}})
+f_{10}(h^{{\beta}}{}_a{\tilde R}^{{\alpha}b} + h^{\alpha}{}_c g^{bc}R^{{\beta}}{}_a)
\nonumber\\
+ 2f_{11}h^{{\alpha}}{}_c g^{bc}{\tilde R}_a{}^{{\beta}}+ 2f_{12}h^{{\alpha}}{}_c g^{bc}
{\tilde R}^{{\beta}}{}_a + f_{13}({\delta}^b_aR^{{\beta}{\alpha}} + h^{{\beta}}{}_aV^{b{\alpha}})
\nonumber\\
+ f_{14}({\delta}^b_a {\tilde R}^{{\beta}{\alpha}}+h^{{\alpha}}{}_c g^{bc}V_a{}^{{\beta}})+
2f_{15}{\delta}^b_aV^{{\beta}{\alpha}}\Bigr)\,T^{\mu}{}_{{\alpha}{\beta}}
\nonumber\\
+2\sqrt{-g}\beta^2\Bigl(2a_1 T_a{}^{\mu b} + 2a_2 T^{[b\mu ]}{}_a + 2a_3\delta_a^{[b}T^{\mu ]}+2k_1Q_a{}^{b\mu}+2k_2Q^\mu{}_{(a}{}^{b)}
\nonumber\\
+2k_3\delta^b_aQ^\mu+2k_4 h^\mu{}_{(a}Q^{b)\nu}{}_\nu  +k_5(h^{\mu}{}_{(a}Q^{b)}+\delta^b_aQ^{\mu\nu}{}_{\nu})
\nonumber\\
+m_1(T_{(a}{}^{b)\mu} - Q_a{}^{[b\mu]})+ m_2(\delta^b_aT^{\mu} +\delta^{[b}_a Q^{\mu]})
+m_3(T_{(a}h^{b)}{}_{\sigma}g^{\sigma\mu}
\nonumber\\
+\delta^{[b}_a Q^{\mu]\lambda}{}_\lambda ) \Bigr)+\sqrt{-g}\beta\partial_\nu\beta \Bigl( 4f_0 h^\mu{}_{[a}h^\nu{}_{c]} g^{bc} + (l_2 + 2l_3)\delta^b_ag^{\nu\mu} 
\nonumber\\
-l_2h^\mu{}_a h^\nu{}_c g^{cb}+2l_4h^\nu{}_{(a}g^{b)c}h^\mu{}_c \Bigr)+
\sqrt{-g}\beta^4\Lambda^{\mu b}{}_a + \frac{\delta\mathcal{L}_m}{\delta\Gamma^a{}_{b\mu}} = 0\, .
\label{eq:vGbeta}
\end{eqnarray}
Then perform the same variational procedure for the tetrad coefficients $h^a{}_\mu$ and obtain a variational $h$-equation,
\begin{eqnarray}
\fl
\frac{\delta\mathcal L}{\delta h^a{}_\mu}=
\sqrt{-g}\biggl[h^\mu{}_a \Bigl(f_1R^{\alpha\beta\sigma\nu}R_{\alpha\beta\sigma\nu}+
f_2R^{\alpha\beta\sigma\nu}R_{\beta\alpha\sigma\nu}+ f_3R^{\alpha\beta\sigma\nu}R_{\alpha\sigma\beta\nu}
\nonumber\\+
f_4R^{\alpha\beta\sigma\nu}R_{\beta\sigma\alpha\nu}+
f_5R^{\alpha\beta\sigma\nu}R_{\sigma\nu\alpha\beta}+f_6R^2+ f_7R^{\sigma\nu}R_{\sigma\nu}
\nonumber\\+
f_8R^{\sigma\nu}R_{\nu\sigma}+
f_9R^{\sigma\nu}\tilde{R}_{\sigma\nu}+ f_{10}R^{\sigma\nu}\tilde{R}_{\nu\sigma}+
f_{11}\tilde{R}^{\sigma\nu}\tilde{R}_{\sigma\nu}
\nonumber\\+
f_{12}\tilde{R}^{\sigma\nu}\tilde{R}_{\nu\sigma}+f_{13}V^{\sigma\nu}R_{\sigma\nu}+
f_{14}V^{\sigma\nu}\tilde{R}_{\sigma\nu}+f_{15}V^{\sigma\nu}V_{\sigma\nu}\Bigr )
\nonumber\\-
4f_1R^{\alpha\beta\mu\nu}R_{\alpha\beta a\nu}-
4f_2R^{\alpha\beta\mu\nu}R_{\beta\alpha a\nu}+2f_3(R^{\alpha\mu\beta\nu}R_{\alpha\beta \nu a}
\nonumber\\+
R^{\alpha\beta\mu\nu}R_{\alpha\nu \beta a}) -f_4(R^{\mu\beta\alpha\nu}R_{\beta \alpha a\nu }+ R^{\beta\mu\alpha\nu}R_{\alpha\beta a\nu }
\nonumber\\+
R^{\alpha\beta\nu\mu}R_{\beta\nu\alpha a}+ R^{\beta\nu\alpha\mu}R_{\alpha\beta\nu a})+4f_5R^{[\mu\nu]\alpha\beta}R_{\alpha\beta \nu a}
\nonumber\\-
2f_6R(\tilde{R}^{\mu}{}_a + R^{\mu}{}_a)-2f_7(R^{\sigma\mu}R_{\sigma a}+R^{\sigma\nu}R^{\mu}{}_{\sigma a\nu})
\nonumber\\-
2f_8(R^{\mu\nu}R_{\nu a}
+R^{\mu \sigma}{}_a{}^{\nu}R_{\nu \sigma})-f_9(R^{\mu\sigma}{}_a{}^{\nu}\tilde{R}_{\sigma \nu}+ R_{\sigma\nu}R^{\sigma\mu\nu}{}_a
\nonumber\\+
{R}_{\nu a}\tilde{R}^{\nu\mu} + R^{\nu\mu}\tilde{R}_{\nu a})-f_{10}(R^{\mu\nu}{}_a{}^\sigma\tilde{R}_{\sigma\nu} + R^{\mu\nu}\tilde{R}_{\nu a}
\nonumber\\+
R^{\sigma\mu\nu}{}_a R_{\nu \sigma} + R_{\nu a}\tilde{R}^{\mu\nu})-2f_{11}(R^{\sigma\mu\nu}{}_a\tilde{R}_{\sigma\nu}+\tilde{R}^{\nu\mu}\tilde{R}_{\nu a})
\nonumber\\-
2f_{12}(R^{\sigma\mu\nu}{}_a\tilde{R}_{\nu \sigma} + \tilde{R}^{\mu \sigma}\tilde{R}_{\sigma a})
-f_{13}(2V^{\mu \nu}R_{[a\nu]}+2V_{a\nu}R^{[\mu \nu]}
\nonumber\\+
V^{\sigma\nu}R^\mu{}_{\sigma a\nu}-V^{\mu \nu}R_{a\nu})-f_{14}(2V^{\sigma(\mu}\tilde{R}_{|\sigma|a)}+V_{\sigma\nu}R^{\sigma\mu\nu}{}_a
\nonumber\\+
V_{a\nu}\tilde{R}^{\mu\nu})-4f_{15}V^{\mu\nu}V_{a\nu}\biggr ]
\nonumber\\
+\sqrt{-g}\beta^2\biggl[h^\mu{}_a\Bigl(a_1T^{\alpha\sigma\nu}T_{\alpha\sigma\nu}+a_2T^{\alpha\sigma\nu}T_{\nu\sigma\alpha}+a_3 T^{\nu}T_{\nu}
\nonumber\\+
k_1Q^{\sigma\nu\lambda}Q_{\sigma\nu\lambda}+k_2Q^{\sigma\nu\lambda}Q_{\sigma\lambda\nu}+ k_3Q^\nu Q_\nu+k_4Q_{\lambda}{}^{\sigma}{}_{\sigma}Q^{{\lambda}{\nu}}{}_{\nu}
\nonumber\\+
k_5Q^\sigma Q_{\sigma}{}^\nu{}_{\nu}+m_1 Q^{\sigma{\nu}{\lambda}}T_{\sigma{\nu}{\lambda}}+ m_2Q^\nu T_\nu+ m_3Q^\sigma {}_{{\nu}\sigma}T^{\nu}\Bigr)
\nonumber\\
-f_0(R^{\mu}{}_a+\tilde{R}^{\mu}{}_a-h^\mu{}_a R)+2a_1(T^{\mu\sigma\nu}T_{a\sigma\nu} - 2T^{\sigma\nu\mu}T_{\sigma \nu a})
\nonumber\\-
2a_2T^{\nu\mu\sigma}T_{\sigma a\nu}-2a_3T^{\mu} T_{a}-2k_1 Q^{\nu\lambda\mu}Q_{\nu\lambda a}-2k_2 Q^{\mu\nu\lambda}Q_{\nu\lambda a}
\nonumber\\-
2k_3 Q^\mu Q_a-2k_4Q^{\lambda\mu }{}_a Q_\lambda{}^\nu{}_\nu - k_5(Q^{\mu\nu}{}_\nu Q_a + Q_\nu Q^{\nu\mu}{}_a )
\nonumber\\
+ m_1(T^{\mu\lambda\nu}Q_{a\lambda\nu} - Q^{\mu\nu\lambda}T_{\nu a\lambda}
-T^{\nu\lambda\mu}Q_{\nu\lambda a}-Q^{\nu\lambda\mu}T_{\nu\lambda a})
\nonumber\\
- m_2(Q^\mu T_a +T^\mu Q_a)-m_3 (Q^{\mu\nu}{}_a T_\nu +Q^{\mu\nu}{}_\nu T_a)\biggr]
\nonumber\\
+\sqrt{-g}\biggl[h^\mu{}_a\Bigl(l_1g^{\sigma\rho}\partial_\sigma\beta\partial_\rho\beta + l_2\beta T^\nu\partial_\nu\beta + l_3\beta Q^\nu\partial_\nu\beta 
\nonumber\\
+ l_4\beta Q^{\nu\sigma}{}_\sigma\partial_\nu\beta + \Lambda_0\beta^4\Bigr) - 2l_1\partial_\sigma\beta\partial_\rho\beta g^{\mu\sigma}h^\rho{}_a 
\nonumber\\
-2\beta\partial_\nu\beta(l_2T_\sigma g^{\mu (\sigma} + l_3Q_\sigma g^{\mu (\sigma} + l_4Q^{\mu (\sigma}{}_\sigma ) h^{\nu)}{}_a\biggr]
\nonumber\\
+2\stackrel{*}{\nabla}_\nu\biggl[\sqrt{-g}\beta^2\Bigl(2a_1T_a{}^{\mu\nu} + 2a_2T^{[\nu\mu]}{}_a + 2a_3h^{[\nu}{}_{|a|}T^{\mu]}
\nonumber\\
+m_1Q_a{}^{[\mu\nu]} - m_2h^{[\mu}{}_a Q^{\nu]} - m_3h^{[\mu}{}_a Q^{\nu]\lambda}{}_\lambda + l_2\partial_\sigma(\ln\beta)g^{\sigma[\mu}h^{\nu]}{}_a\Bigr)\biggr]
\nonumber\\
+\sqrt{-g}\beta^4\Lambda^{\sigma}{}_{bc}Q^{bc}{}_\sigma h^\mu{}_a + \frac{\delta\mathcal{L}_m}{\delta h^a{}_{\mu}} = 0\, .
\label{eq:vhbeta}
\end{eqnarray}

In the approach developed, the additional equation arises as a result of the variation of the Dirac's scalar field $\beta$. The corresponding variational equation ($\beta$-equation) reads,
\begin{eqnarray}
\fl
\frac{\delta\mathcal{L}}{\delta\beta} = - 2l_1\stackrel{*}{\nabla}_\mu\left(\sqrt{-g}g^{\mu\nu}\partial_\nu\beta\right) - \beta\stackrel{*}{\nabla}_\mu\left(\sqrt{-g}(l_2T^\mu + l_3Q^\mu + l_4Q^{\mu\lambda}{}_\lambda)\right)
\nonumber\\
+2\beta\sqrt{-g}\biggl(f_0R +a_1T^{{\lambda}{\mu}{\nu}}T_{{\lambda}{\mu}{\nu}}+
a_2T^{{\lambda}{\mu}{\nu}}T_{{\nu}{\mu}{\lambda}}+ a_3T^\mu T_\mu 
\nonumber\\
+k_1Q^{{\mu}{\nu}{\lambda}}Q_{{\mu}{\nu}{\lambda}}+ k_2Q^{{\mu}{\nu}{\lambda}}Q_{{\mu}{\lambda}{\nu}}+k_3Q^\mu Q_\mu +
k_4Q_{\lambda}{}^{\mu}{}_{\mu}Q^{{\lambda}{\nu}}{}_{\nu}
\nonumber\\
+k_5Q^\mu Q_{\mu}{}^\nu{}_{\nu}
+m_1Q^{{\mu}{\nu}{\lambda}}T_{{\mu}{\nu}{\lambda}}+
m_2Q^{\mu}T_{\mu}+ m_3Q^{\mu}{}_{{\nu}{\mu}}T^{\nu}\biggr) 
\nonumber\\
+ 4\sqrt{-g}\beta^3 (\Lambda_0 + \frac{1}{2}\Lambda^\mu{}_{ab}Q^{ab}{}_\mu) +
\frac{\delta \mathcal{L}_m}{\delta\beta} = 0\,.
\label{eq:betaur}
\end{eqnarray}

As it has been pointed out at the end of section 3, we can formally write down the result of variation of  Lagrangian density (\ref{eq:Lagr}) under the components of the metric tensor $g^{ab}$. The corresponding $g$-equation reads, 
\begin{eqnarray}
\fl
\frac{\delta\mathcal L}{\delta g^{ab}} = \sqrt{-g}f_0\left(R_{(ab)}-\frac12g_{ab}R\right)\nonumber\\
-\frac12\sqrt{-g}g_{ab}\Bigl(f_1R^{{\alpha}{\beta}{\mu}{\nu}}R_{{\alpha}{\beta}{\mu}{\nu}} +
f_2R^{{\alpha}{\beta}{\mu}{\nu}}R_{{\beta}{\alpha}{\mu}{\nu}}  \nonumber\\
 +f_3R^{{\alpha}{\beta}{\mu}{\nu}}R_{{\alpha}{\mu}{\beta}{\nu}}+
f_4R^{{\alpha}{\beta}{\mu}{\nu}}R_{{\beta}{\mu}{\alpha}{\nu}}+
f_5R^{{\alpha}{\beta}{\mu}{\nu}}R_{{\mu}{\nu}{\alpha}{\beta}}\nonumber\\
+ f_6R^2+f_7R^{{\mu}{\nu}}R_{{\mu}{\nu}}+
f_8R^{{\mu}{\nu}}R_{{\nu}{\mu}} + f_9R^{{\mu}{\nu}}{\tilde R}_{{\mu}{\nu}} \nonumber\\
+ f_{10}R^{{\mu}{\nu}}{\tilde R}_{{\nu}{\mu}}+ f_{11}{\tilde R}^{{\mu}{\nu}}{\tilde
R}_{{\mu}{\nu}}+ f_{12}{\tilde R}^{{\mu}{\nu}}{\tilde R}_{{\nu}{\mu}} \nonumber\\
+f_{13}V^{{\mu}{\nu}}R_{{\mu}{\nu}}+ f_{14}V^{{\mu}{\nu}}{\tilde
R}_{{\mu}{\nu}}+f_{15}V^{{\mu}{\nu}}V_{{\mu}{\nu}}\nonumber\\
 +\beta^2 (a_1T^{\lambda\mu\nu}T_{\lambda\mu\nu }+a_2T^{\lambda\mu\nu}T_{\nu\mu\lambda}
+ a_3T^{\mu}T_{\mu} \nonumber\\
+k_1Q^{{\mu}{\nu}{\lambda}}Q_{{\mu}{\nu}{\lambda}}+
k_2Q^{{\mu}{\nu}{\lambda}}Q_{{\mu}{\lambda}{\nu}}+k_3Q^\mu Q_\mu  \nonumber\\
+k_4Q_{\lambda}{}^{\mu}{}_{\mu}Q^{\lambda\nu}{}_{\nu}+
k_5Q^\mu Q_{\mu}{}^\nu{}_{\nu} \nonumber\\
+m_1Q^{{\mu}{\nu}{\lambda}}T_{{\mu}{\nu}{\lambda}}+
m_2Q^{\mu}T_{\mu}+ m_3Q^{\mu}{}_{{\nu}{\mu}}T^{\nu}) \nonumber\\
+l_1g^{\mu\nu}\partial_\mu\beta\partial_\nu\beta + l_2\beta\partial_\mu\beta g^{\mu\sigma} T_\sigma 
+ l_3\beta\partial_\mu\beta g^{\mu\sigma}Q_\sigma  \nonumber\\
+ l_4\beta \partial_\mu\beta Q^{\mu\sigma}{}_{\sigma} + \Lambda_0 \beta^4\Bigr)\nonumber\\
+\sqrt{-g}\Bigl(-f_1(R_{(a|\sigma\mu\nu |}R_{b)}{}^{\sigma\mu{\nu}}-R^{\sigma}{}_{(a}{}^{\mu{\nu}}
R_{|\sigma |b)\mu{\nu}}\nonumber\\
+2R_{\sigma\mu\nu(a}R^{\sigma\mu}{}_{b)}{}^{\nu})+2f_2R^{\sigma\mu}{}_{(a}{}^{\nu}R_{|\mu\sigma | b){\nu}} \nonumber\\
+ f_3(R_{\sigma\mu{\nu}(a}R^{\sigma{\nu}\mu}{}_{b)}-
2R_{\sigma\mu\nu (a}R^\sigma{}_{b)}{}^{\mu{\nu}}- R_{(a|\sigma\mu{\nu}|}R_{b)}{}^{\mu\sigma{\nu}})\nonumber\\
+ f_4(-R^{\mu\sigma\nu}{}_{(a}R_{|\sigma |b)\mu{\nu}}+R^{{\nu}\mu\sigma}{}_{(a} R_{|\mu\sigma{\nu}|b)})\nonumber\\
+ 2f_5R^{\sigma\mu{\nu}}{}_{(a}R_{|{\nu}|b)\sigma\mu}+ 2f_6 RR_{(ab)}\nonumber\\
+f_7(R_{(a}{}^\sigma R_{b)\sigma}+R^\sigma{}_{(a}R_{|\sigma |b)})
+2f_8R_{(a}{}^\sigma R_{|\sigma |b)} \nonumber\\
+ f_9(R_{\mu (a}{\tilde R}^\mu{}_{b)} - R^{\mu\nu}R_{\mu (ab)\nu})
+ f_{10}({\tilde R}^\mu{}_{(a}R_{b)\mu}-R^{\mu\nu}R_{\nu(ab)\mu}) \nonumber\\
+f_{11}(-2R_{\mu (ab)\nu}{\tilde R}^{\mu\nu}-{\tilde R}_{a \nu }{\tilde R}_b{}^\nu
+{\tilde R}_{\nu a}{\tilde R}^\nu{}_b) \nonumber\\
- 2f_{12}{\tilde R}^{\nu\mu }R_{\mu(ab)\nu}+ 2f_{13}V_{(a}{}^\sigma R_{[b)\sigma ]} \nonumber\\
 +f_{14}(V_{\sigma (a}{\tilde R}^\sigma{}_{b)}-V^{\mu\nu}R_{\mu(ab)\nu})
+2f_{15}V_{(a}{}^\nu V_{b)\nu}\Bigr) \nonumber\\
+\sqrt{-g}\beta^2 \Bigl (a_1(2T_{\mu\nu a}T^{\mu\nu }{}_b - T_a{}^{\mu\nu }T_{b\mu\nu })+ 
 a_2T_{\mu\nu (a}T^{|\nu\mu |}{}_{b)} \nonumber\\
+a_3T_aT_b+ k_1(Q^{\mu\nu}{}_aQ_{\mu\nu b} - 2 Q_a{}^{\mu\nu}Q_{b\mu\nu}) - 
k_2 Q_{(a}{}^{\mu\nu}Q_{b)\nu\mu}  \nonumber\\
 +k_3(Q_aQ_b - 2Q_{ab\mu} Q^\mu )-k_4Q_a{}^\mu{}_\mu Q_b{}^\nu{}_\nu  -k_5Q_{ab\mu}Q^{\mu\nu}{}_\nu \nonumber\\
+ m_1 (Q_{\nu\lambda (a}T^{\nu\lambda}{}_{b)}-Q_{(a}{}^{\mu\nu}T_{b)\mu\nu})+
m_2 (T_{(a}Q_{b)} - Q_{ab\mu}T^\mu )\Bigr) \nonumber\\
+\sqrt{-g}\Bigl(l_1h^\mu{}_{(a}h^\nu{}_{b)}\partial_\mu \beta \partial_\nu \beta 
+ (l_2T_{(a}h^\nu{}_{b)} + l_3Q_{(a}h^\nu{}_{b)} - l_3Q_{ab}{}^\nu )  \beta\partial_\nu\beta\Bigr)\nonumber\\
-\stackrel{*}{\nabla}_\mu\Bigl(\sqrt{-g}\beta^2 (2k_1Q_{ab}{}^\mu + 2k_2Q^\mu{}_{(ab)} 
+ 2k_3g_{ab}Q^\mu \nonumber\\
+ 2k_4h^\mu{}_{(a}Q_{b)}{}^\nu {}_\nu + k_5 (g_{ab}Q^{\mu\nu}{}_\nu +Q_{(a} h^\mu{}_{b)})\nonumber\\
+ m_1T_{(ab)}{}^\mu + m_2g_{ab}T^\mu + m_3T_{(a}h^\mu{}_{b)}\nonumber\\
+(l_3g_{ab}g^{\mu\nu} + l_4h^\mu{}_{(a}h^\nu{}_{b)})\partial_\nu\ln\beta )\Bigr)\nonumber\\
-\frac12\sqrt{-g}\beta^4\Bigl(\hat\nabla_\mu\Lambda^\mu{}_{ab} + 2\Lambda^\mu{}_{cd}Q^{cd}{}_\mu g_{ab}\Bigr) + \frac{\delta\mathcal L_m}{\delta g^{ab}} = 0\,.
\label{eq:vg}
\end{eqnarray}
In this equation the Weyl condition (\ref{eq:Cond2}) should be substituted.


\section*{References}
\begin{thebibliography}{99}
\bibitem{BFKRusGrav13}
Babourova O V, Kostkin R S and Frolov B N 2009 {\it Proc. 13 Russian Gravitational Conference RUSGRAV-13} (Moscow: PFUR) p~25 (in Russian)
\bibitem{BFKIzvVuz2008}
Baburova O V and Kostkin R S 2009 {\it Russian Physics Journal} {\bf 52} p~487
\bibitem{BFZ1}
Babourova O V, Frolov B N and Zhukovsky V Ch 2006 \PR {\it D} {\bf 74} p~064012 ({\it Preprint} gr-qc/0508088)
\bibitem{BFZ2}
Baburova O V, Zhukovsky V Ch and Frolov B N 2008 {\it Theoretical and Mathematical Physics} {\bf 157} p 1420
\bibitem{BFZ3}
Babourova O V, Frolov B N and Zhukovsky V Ch 2009 {\it Gravitation and Cosmology} {\bf 15} p 13
\bibitem{Weyl2}
Weyl H 1952 {\em Space, time, matter} (New York: Dover)
\bibitem{Fr1}
Frolov B N 1963 in {\it Vestnik Mosk. Universiteta. Ser. 3: Fiz., Astron.} No {\bf 6} p~48 (in Russian)
\bibitem{Fr2}
Frolov B N  1967 in {\it Modern Problems of Gravitation, Proc. 2-nd Soviet Gravit. Conf.} (Tbilisi) (in Russian)
\bibitem{Fr:book}
Frolov B N 2003 {Poincar\'{e}-gauge theory of gravitation} (Moscow: Moscow State Pedagogical University Press) (in Russian)
\bibitem{FrIv}
Frolov B N 2004 {\it Gravitation and Cosmology} {\bf 6} p~116
\bibitem{Ut2}
Utiyama R 1973 {\it Progress of Theor. Phys.} {\bf 50} p~2080
\bibitem{Freud}
Freud P G O 1974 \APNY {\bf 84} p~440
\bibitem{Ut3}
Utiyama R 1975 {\it Gen. Rel. Grav.} {\bf 6} p~41
\bibitem{Dir}
Dirac P A M 1973 \PRS {\bf A333} p~403
\bibitem{Frolov4}
Frolov B. N. 1996 in {\em Gravity, Particles and spacetime} (Ed. P. Pronin and G.
Sardanashvily) {Singapore, New Jersey, London, Hong Kong: World Scientific} P~113 
\bibitem{Duff}
Duff M J 1994 \CQG {\bf 11} p~1387 ({\it Preprint} hep-th/9308075)
\bibitem{Stromg} Strominger A 1990 \NP {\bf B343} p~167
\bibitem{BFKGRACOS2009}Babourova O V, Kostkin R S and Frolov B N 2009 {\it Proc. II Russian summer school-seminar -- Modern theoretical problems of gravitation and cosmology GRACOS-2009} (Russia: Kazan-Yalchik) p~127 (in Russian)
\bibitem{BFK_lanl} Babourova O V, Frolov B N and Kostkin R S 2010 Dirac's scalar field as dark energy within the frameworks of conformal theory of gravitation in Weyl--Cartan space ({\it Preprint}  gr-qc/1006.4761) 
\bibitem{BFK_Izv} Babourova O V, Kostkin R S, Frolov B N 2011 {\it Izv.Vuz.Fiz.} {\bf 54} N 1 p~109 (in Russian)
\bibitem{Gre-Pap1} Gregorash D and Papini G 1980 {\it Nuovo Cim} {\bf 55B} p~37
\bibitem{Gre-Pap2} Gregorash D and Papini G 1980 {\it Nuovo Cim} {\bf 56B} p~21
\bibitem{Perv1} Pervushin V, Proskurin D 2002 {\it Gravitation and Cosmology} {\bf 8} p~161 ({\it Preprint} gr-qc/0106006)
\bibitem{Behnke} Behnke D, Blaschke D, Pervushin V N, Proskurin D 2002 {\it Phys. Lett. B}  {\bf 530} P~20
\bibitem{Glinka} Glinka L A, Pervushin V N  2008 {\it Old New Concepts Phys} {\bf 5} P~31 ({\it Preprint} gr-qc/0705.0655)
\bibitem{Arbuzov} Arbuzov A B et al 2010 Conformal Hamiltonian Dynamics of General Relativity ({\it Preprint} gr-qc/1007.0293)
\bibitem{IntJ} Babourova O V and Frolov B N 1997 {\it Int. J. Mod. Phys. A} {\bf 12} p~3665 ({\it Preprint} gr-qc/9609004)
\bibitem{ModPhysLet} Babourova O V and Frolov B N 1997 {Mod. Phys. Lett. A} {\bf 12} p~1267 ({\it Preprint} gr-qc/9609005)
\bibitem{MinkGarKud} Minkevich A V, Garkun A S and Kudin V I 2007 \CQG {\bf 24} p~5835 ({\it Preprint} gr-qc/0706.1157).
\bibitem{BKU} Baburova O V, Korolev V F and Umyarova I Ya  2006 {\it Russian Physics Journal} {\bf 49} p~531
\bibitem{BabK:izv} Babourova O V  and Korolev V F 2006 {\it Russian Physics Journal} {\bf 49} p~628
\bibitem{Schr} Schr\"{o}dinger E 1950 {\it spacetime structure} (Cambridge: Cambridge Univ. Press)
\bibitem{Fer-FR} Ferraris M, Francaviglia M and Reina C 1982 {\it Gen. Rel. Grav} {\bf 14} p~243
\bibitem{Fr3} Frolov B N 1978 {\it Acta Phys. Polon.} {\bf B9} p~823
\bibitem{Tsamp} Tsamparlis M 1979 \PL {\bf 75A} p~27
\bibitem{Min1} Minkevich A V 1980 \PL {\bf 80A} p~232
\bibitem{CQG} Baburova O V, Frolov B N and Klimova E A  2003 \CQG {\bf 20} p~1423
\bibitem{Fock} Fock V 1964 {\it The theory of space, time and gravitation} (New-York: Macmillan in New-York)
\bibitem{Mink2} Minkevich A V 2009 \PL {\it B} {\bf 678} p~423 ({\it Preprint} gr-qc/0902.2860v2).
\bibitem{FrKouch} Frolov B N, Kouchoumov A Ju 2009 In {\em Physical Interpretations of Relativity Theory (Proc. XV Int. Sci. Meeting PIRT-2009)} (Moscow: Bauman Moscow State Technical University) P~359 
\bibitem{Tr2} 
Trautman A 1972 {\em Bul. Acad. Pol. Sci. (Ser. sci. math., astr., phys.} {\bf 20} No 6 p~503  
\bibitem{YadPhys2}
Babourova O V,  Vshivtsev A S, Myasnikov V P, Frolov B N 1998 {\em Phys. Atom. Nucl.} {\bf 61} p~2175  ({\em Preprint} hep-th/0407153)
\bibitem{BabFrKor}  
Baburova O V, Korolev M Ju and Frolov B N 1994 {\it Russian Physics Journal} {\bf 37} No~1  p~76  
\bibitem{He-CMN}
Hehl F W, McCrea J L, Mielke E W and Ne\'{e}man Yu 1995 {\em Phys. Rep.} {\bf 258} p~1 ({\em Preprint} gr-qc/9402012)
\bibitem{DilSing2}
Babourova O V and Frolov B N 1998 {\em Mod. Phys. Lett. A} {\bf 13} p~7 ({\em Preprint} gr-qc/9708009)
\bibitem{BFKgrandkosm2009}
Babourova O V, Kostkin R S, Frolov B N 2009 {/em Gravation and Cosmology} {\bf 15} No 4 p~302
\bibitem{KostIzvVuz2009} 
Kostkin R S 2009 {\it Russian Physics Journal} {\bf 52} No 9 P~98
\bibitem{BFKPIRT2009}
Babourova O V, Frolov B N, Kostkin R S in {\em Physical Interpretations of Relativity Theory (Proc. XV Int. Sci. Meeting PIRT-2009)} (Moscow: Bauman Moscow State Technical University) P~384
\bibitem{Soleng}
Soleng H H 1996 in {\em Relativity and Scientific Computing -- Computer Algebra, Numerics, Visualization},
(Berlin: Springer) p~210
\bibitem{CosmConPr} Weinberg S 1989 \RMP {\bf 61} p~1
\bibitem{Rubakov} Gorbunov D S, Rubakov V A 2011 {\it Introduction to the Theory of the Early Universe: Hot Big Bang Theory} (Singapore: World Scientific Publ. Co.) in press

\end{thebibliography}
\end{document}